# Giant coercivity enhancement and dimensional crossover of superconductivity in $Co_2FeSi$-NbN nanoscale bilayers

Anurag Gupta[1], Gyanendra Singh[1], Dushyant Kumar[2], Hari Kishan[1] and R. C. Budhani[1,2,*]

[1]National Physical Laboratory (CSIR), Dr. K.S.Krishnan Road, New Delhi-110012 India
[2]Condensed Matter - Low Dimensional Systems Laboratory, Department of Physics, Indian Institute of Technology Kanpur, Kanpur - 208016, India

Proximity coupling and magnetic switching dynamics are shown to be correlated in bilayers (B) of fully spin polarized ferromagnet (F) $Co_2FeSi$ and superconductor (S) NbN. The upper critical field derived from resistivity measurements shows a dimensional crossover with a reduced effective thickness of the S layer. At temperatures (T) << superconducting $T_c$, the measured M-H loops show two step switching; one at T-independent value ~ 7 Oe and other at strongly T-dependent value becoming very large ~ 1 kOe at 2 K. These results reveal induced ferromagnetism in S-layer at the S/F interface with vortex pinning influenced dynamics.

The research field of superconducting (S) and ferromagnetic (F) bilayers (FSB), trilayers and multilayers has gained tremendous importance[1] due to their potential for magnetoeletronics related applications[2] ranging from spin valve to Josephson junctions based devices. The two macroscopic quantum states in S and F layers influence each other via the penetration of superconducting (proximity effect)[2] and ferromagnetic (inverse proximity effect)[3-5] order through their common interface. In presence of strong enough magnetic fields, the vortices in the S-layer can couple with the magnetic domains of the F-layer leading to interesting effects of domain wall pinning of the vortices and vice-a-versa, and domain wall superconductivity[6-10]. Most of such studies have been carried out with ferromagnetic component consisting of 3d transition metals like Fe, Co and Ni, and their alloys. Bilayers of strongly type II superconductors with fully spin polarized ferromagnetic (FSPF) Heusler alloys, which ideally have only one spin state and can potentially show fascinating electronic properties, have not been studied yet. Recently, our group has explored the use of the Co-based Heusler compound ($Co_2MnSi$) electrodes for efficient injection of pure spin currents into superconducting NbN[11]. NbN is a type II superconductor of short coherence length ($\xi(0) \approx 5$ nm) and long penetration depth ($\lambda(0) \approx 200$ nm) while $Co_2FeSi$ (CFS) is a ferromagnet of the Heusler family with high magnetic ordering temperature ~ 1150 K and saturation magnetization 6 $\mu_B$ per formula unit[12]. In the present work we investigate the nature of proximity coupling and its influence on magnetic switching dynamics in the hitherto unexplored $NbN/Co_2FeSi$ bilayers grown on (100) MgO.

Thin film bilayers of superconducting NbN and CFS were deposited by pulsed laser ablation (PLD). The 35 nm thick NbN film first deposited on MgO at $200^0$ C in nitrogen environment by ablating 99.99 % Nb target and then CFS layer of 35 nm thick was grown at the same temperature from a stoichiometric CFS target. To establish the magnetic ordering in CFS film, the FSB sample was annealed at 500 $^0$C for 1 hour. For comparison, a pure NbN film was also grown during the same PLD run. The other growth parameters for NbN and $Co_2FeSi$ layers can be found in our previous reports[13,14]. The crystalline quality and interface roughness of the films were analyzed using X-ray diffraction and reflectivity respectively. The actual thickness of each layer was estimated from the best fitting of X-ray reflectivity data. Magnetization and electrical transport measurements as a function of temperature (2 - 20 K) and magnetic field (0 - 7 T) were carried in a Quantum Design MPMS system.

Fig.1(a) shows X-ray diffraction pattern of $NbN/Co_2FeSi$ bilayer thin film. We observed sharp (002) and (004) reflections for the NbN and $Co_2FeSi$, indicating highly (00l) textured growth of both the materials. The small lattice mismatch between NbN ($á \approx 4.38$ Å) and MgO substrate ($a \approx 4.21$ Å) should promote epitaxial (cube on cube) growth of the film. However, in order to maintain the epitaxiallity of CFS (($a \approx 5.65$ Å) on NbN/MgO plane, rotation of the unit cell by $45^o$ (about [001] direction) with respect to substrate and NbN is required. A sketch of the basal plane of unit cell for CFS, NbN and MgO is shown in Fig. 1(b). The difference between $a$ and $á\sqrt{2}$ is only 8%. The X-ray reflectivity shows interface roughness of ~ 1.5 nm between NbN and CFS layer. The resistively measured $T_c$ of the pure NbN and FSB samples as shown in Fig. 1(c) are ~ 13.3 and 11.8 K. These values are in good agreement with ~13.1 and 11.4 K measured magnetically with an applied in-plane H = 50 Oe, as shown in Fig. 1(d). The 1.4 K decrease in the $T_C$ of FSB indicates the influence of F-layer (CFS) on the superconductivity of NbN layer. Note that in the FSB sample, unlike for the pure NbN film, the value of M at T > $T_c$ is positive ~ 4 x $10^{-4}$ emu and not zero. This indicates that the ferromagnetic signal of CFS layer is superimposed on the superconducting signal due to NbN layer in the FSB.

To look at the magnetic response of FSB and pure NbN samples, we first show in Fig. 2 the characteristic H-T phase diagram measured both magnetically and resistively. The H*(T) curves in the Fig.2 correspond to the characteristic T, at different applied in-plane constant $H_{//}$, where both FC and ZFC M(T) merge together. In case of both the samples the H*(T) curve lies very low in comparison to their observed $H_{c2}$(T) curve, reflecting a drastic decrease in the irreversible region of the superconducting state in the H-T phase diagram. One possible reason could be the much smaller thickness of the NbN-layer ~ 35 nm in both the samples compared to the typical penetration depth in NbN ($\lambda$(0) ~ 200 nm), that can lead to a suppressed irreversible behavior of the S-layer. In contrast, the upper critical field $H_{c2}$(T) of both the samples, determined resistively at different applied in-plane constant $H_{//}$, lies very high in the H-T phase diagram (see Fig. 2). For both pure NbN and FSB, the observed values near $T_c$, of $dH_{c2}$(T)/dT ~ 35 and 30 kOe/K are much higher than dH*/dT = 1.79 and 1.75 kOe/K, respectively. In addition, for both the samples $H_{c2}$(T) shows a linear behavior at T << $T_c$ and a nonlinear behavior at T < $T_c$. We identify this result with a 2D-3D crossover, occurring at a characteristic temperature $T_{CR}$, which has been frequently observed in S and S-F heterostructures[15-17]. At lower temperatures, when the coherence length perpendicular to the plane of the film $\xi_\perp$(T) < $d_s$ (S-layer thickness) the film behaves like a 3D system. While close to $T_C$ where $\xi_\perp$(T) diverges, $\xi_\perp$(T) > $d_s$ and the film shows a 2D behavior. Although, NbN is an s-wave isotropic superconductor as a bulk, in thin film ($d_s$ << $\lambda$) form it may show a geometrical anisotropy, i.e. $H_{c2//}$(T) and $H_{c2\perp}$(T) can be different[18]. The NbN thin films have been reported to show anisotropic behavior also due to the grain morphology[19]. Thus, we prefer to employ an anisotropic GL (Ginzburg Landau) model to describe the behavior of $H_{c2//}$(T) in our films, which in the 3D regime ($\xi_\perp$(T) < $d_s$) is given by[20,15]:

$$\mu_0 H_{c2||}(T) = \frac{\phi_0}{2\pi\xi_{||}(0)\xi_\perp(0)}\left(1 - \frac{T}{T_c}\right) \quad (1)$$

whereas in the 2D regime ($\xi_\perp$(T) > $d_s$) is expressed as[20,15]:

$$\mu_0 H_{c2||}(T) = \frac{\sqrt{12}\phi_0}{2\pi\xi_{||}(0)d}\sqrt{1 - \frac{T}{T_c}} \quad (2)$$

where $\xi_{//}$(0) and $\xi_\perp$(0) are the zero temperature coherence length parallel and perpendicular to the plane of superconducting thin film. The fits of the equations 1 & 2 are shown in Fig. 2. We clearly observe crossover behavior with $T_{CR}$ ~ 10.4 K and 12.9 K for NbN/CFS and pure NbN samples, respectively. The calculated $\xi_{//}$(0) and $\xi_\perp$(0) are: $\xi_{//}^{NbN}$(0) ~ 18.5Å, $\xi_\perp^{NbN}$(0) ~ 40.8 Å, $\xi_{//}^{FSB}$(0) ~ 16.3 Å and $\xi_\perp^{FSB}$(0) ~ 73.1 Å. It is interesting to note that the values of $T_C$-$T_{CR}$ for pure NbN and FSB are 0.2 and 1 K respectively. While the thickness of NbN is same in both the samples, the higher value of $T_C$-$T_{CR}$ in FSB sample can be attributed to the reduction of the effective thickness ($d_{eff}$) of the NbN layer in FSB due to the proximity of the CFS layer. Our calculations show that the $d_{eff}$ of the S-layer in pure NbN and FSB at $T_{CR}$ is ~33 and 24 nm, respectively. Note that calculated $d_{eff}$ for pure NbN film is close to the actual thickness (~35 nm) of the film, which confirms the reliability of the data. Similar behavior of dimensional crossover is seen in other systems of FSBs with soft ferromagnet where $T_C$-$T_{CR}$ ~1 K is reported[15].

The observation of a square hysteresis loop with very low coercivity (as shown below) in the normal state (T >$T_c$) suggests that the CFS layer is a single magnetic entity. In that case, the Cooper pair electrons with opposite spins in the NbN layer will not diffuse into a fully spin polarized CFS layer, which makes the origin of the decrease in $T_c$ and $d_{eff}$ in FSB with respect to pure NbN a-priori unclear. There are two other possible scenarios. The first is inverse proximity mechanism[3-5], where the F-layer may not only suppress the superconducting order but may also induce ferromagnetism in the S-layer next to the S/F interface. The thickness of such layer can be ~ 7-10 nm[3,4], which is near to the value ~ 11 nm extracted from 3D-2D crossover. The second scenario is where a p-wave (spin triplet) pairing is proximity induced in the CFS layer, which depletes the S-layer order parameter. The characteristic ferromagnetic coherence length $\xi_F$ follows the relation $\xi_F = \left(\hbar D_F/2\pi K_B T_{CS}\right)^{1/2}$, where $D_F$ is the diffusion coefficient of the ferromagnet. $D_F$ is related to low temperature resistivity ($\rho_F$) and electronic specific heat ($\gamma_F$) by[21] $D_F = \frac{1}{\rho_F \gamma_F}\left(\frac{\pi K_B}{e}\right)^2$. The value of $T_{CS}$ for bilayer is 11.4 K, while the resistivity of $Co_2FeSi$ at 10 K is 70 µΩcm. Taking electronic specific heat coefficient[22] as 2.1 x $10^{-4}$ J/$K^2cm^3$, the coherence length $\xi_F$ is estimated to be 4.2 nm. It is worth mentioning that although both the scenarios can lead to the reduction of $T_c$ in our FSB system, the observed effective reduction of S-layer thickness cannot be understood without invoking the inverse proximity mechanism.

Now we discuss the magnetic response as observed in the isothermal M-H loops measured for both pure NbN and FSB samples. The pure NbN sample shows typical irreversible M-H loops expected for a type II superconductor at T = 4 and 9 K (see Fig. 3a). With increase in T from 4 to 9 K, the significant decrease observed in the M-H loop width is consistent with an expected decrease in vortex pinning strength or the critical current density ($J_c$) of pure NbN. For the FSB sample, the in-plane M-H loops at T = 4 and 12 K are shown in the Fig. 3b and its inset, respectively. At T = 12 K > $T_c$, the M-H loop resembles a typical response of F-layer (CFS) with a coercive field $H_{co}$ ~ 7 G. No contribution from the S-layer (NbN) is discernible. Whereas, at T = 4 K (< $T_C$), the M-H loop of FSB is drastically different from that observed for pure NbN layer or CFS layer. The initial diamagnetic M(H) response observed until H ~ 75 Oe (see the 4$^{th}$ quadrant of the M-H loop in Fig. 3b) reflects the superconductivity of the NbN S-layer. With further increase in H until upto H > 400 Oe the M(H) decreases, followed by a rapid switching from negative to positive value that tends to saturate at ~ 4.3 x $10^{-4}$ emu. This matches exactly with the measured saturated M value ~ 4.3 x $10^{-4}$ emu (at H>1 kOe) at T=12K > $T_c$ (MH shown in the inset of Fig.3b is magnified for clarity). Interestingly, further increasing and later decreasing the field at T=4K leads to an opening of the M-H loop and the value of M rises to ~ 10 x $10^{-4}$ emu until near H ~ 40 Oe. This "paramagnetic hysteresis" may result from the superconducting signal from NbN S-layer superimposed on the magnetic signal of the adjacent CFS F-layer, which will be taken up in details later. Further decrease of H and reversing its direction results in a two step switching in M(H). First a sharp decrease very close to H = 0 in the 2$^{nd}$ quadrant marked as first switching field $H_1$ (~ 7 Oe) followed by a gradual decrease in M(H) is observed. With further increase in reversed field a precipitous drop in magnetization is again observed marked as the second switching field $H_2$ (~ 400 Oe). The value of $H_2$ is same as observed in the 1$^{st}$ quadrant of the M-H loop. In contrast, with field applied out-plane, the behavior of M(H) for FSB sample is totally different. The results at T both below and above $T_c$ are shown in the Fig. 3c and its inset, respectively. At T =20 K (> $T_c$), M(H) shows no saturation even for fields as high as 5 kOe (see inset Fig.3c). At T =4 K (< $T_c$), M(H) shows only a typical type II superconducting response due to the NbN S-layer, with no influence of the CFS F-layer (see Fig.3c). We would conclude that the switching peculiarities of the in-plane M-H behavior of FSB are absent in the out-plane field orientation and pure NbN film.

Previous studies, for instance, Stamopoulos et.al.[23] have shown similar measurements of M-H in NiFe-Nb system just below the $T_c$. However, the unique behavior of two step switching and the giant coercivity observed in our system is different. To see how the switching phenomenon in our FSB system evolves with temperature, we show the in-plane M-H loops at T varying from 2 to 12 K in small steps in Fig.4a and its inset. At T < 7 K, the first switching field $H_1$ is not affected by change in temperature. Whereas, the second switching field $H_2$ is highly T dependent and increase rapidly with decrease in T. At T > 7 K, only a single step switching is visible at $H_1 \sim 7$ Oe, that matches with the $H_{co}$ of the pure CFS layer. Both $H_1$ and $H_2$ are plotted as a function of temperature in Fig.4b. At T<7 K, the simultaneous observation of two step switching at $H_1$ and $H_2$ reveals the evolution of new ferromagnetic region besides the CFS F-layer in our FSB system. As discussed above, a natural candidate for this is the induced ferromagnetism at S/F interface by "inverse proximity" mechanism. However, we would like to discuss two issues here. The first is that the observed ferromagnetism induced in the S-layer has the same sign as that in the F-layer, which is in line with the prediction of ref. [3]. Considering the fact that we have a fully spin polarized ferromagnet CFS as the F-layer, the leakage or tunneling of same sign ferromagnetism into S-layer[3] is more likely rather than the opposite sign induced ferromagnetism[4,5]. The second issue is the observed large value of switching fields $H_2$. Considering the fact that NbN S-layer is a strong type II superconductor, the high values of $H_2$ can result due to the strongly pinned vortices leading to a delay in the magnetic switching of the induced ferromagnetic layer until high fields. Note also that the typical values of $H_2$ lie well below the irreversibility line $H^*(T)$, for instance at 2 K the former (latter) is ~ 1 kOe (20 kOe). Vortex pinning mediated reduction in magnetic domain mobility of S-F bilayers has been reported earlier[9]. The decrease of $H_2$ with increasing T can be understood due to a decrease of vortex pinning in the S-layer, which is the indeed the case follows below.

We finally discuss the observed "paramagnetic hysteresis" of M(H) at H > $H_2$ (see Fig.4a). It reflects the type II superconducting response of the NbN S-layer superimposed on the ferromagnetic signal of the F-layer. With increase in T from 2 to 12 K, the systematic decrease in the width of the M-H loops is a clear signature of decrease in vortex pinning or the $J_c$ of the NbN S-layer. The characteristic T dependent field at which the M-H loop tends to close decreases with T (see Fig.4a) and marks the irreversibility field H*(T) of the NbN S-layer. The observed irreversibility fields are similar to those obtained from FC-ZFC measurements of M(T). Further we plot the loop width ($\Delta M = M(H\uparrow)-M(H\downarrow)$) extracted from the M-H loops (Fig.4a, for H > $H_2$) as a function of T at constant fields in the inset of Fig. 4b. The $\Delta M$ at all H decreases rapidly and vanishes completely around ~ 10-11 K, near the superconducting $T_c(H=0) = 11.4$ K. As seen from the vertical cuts on the x-axis in the inset of Fig.4b, i.e. at different constant T, the value of $\Delta M$ decreases with increasing H. These observations are completely in concurrence with a typical behavior of type II superconductor, and thus $\Delta M(T,H)$ is actually the $J_c(T,H)$ dependence of the NbN S-layer. The increase of vortex pinning with lowering in T in the NbN S-layer gives credence to the T-dependence of $H_2$. The fact that $H_2$ gradually vanishes above ~ 7 K should mean that the vortex pinning at T ≥ 7 K is not strong enough to influence the switching of the induced ferromagnetic layer. It would indeed be interesting to check whether $H_2$ is sample dependent. Preliminary measurements on samples with relatively different strength of the magnetic and superconducting signals do show a variation. The FSB sample chosen in the present work had both the orders equally established with the signals competing with each other. A detailed study of such sample dependence is warranted in future.

In summary, we presented experimental findings on proximity coupling and magnetic switching dynamics of a strong type II superconductor NbN and fully spin polarized ferromagnet $Co_2FeSi$ bilayer. We provided an evidence for "inverse proximity" induced same sign ferromagnetism in the S-layer. The magnetic switching of this induced layer was shown to get delayed to anomalously large fields due to the adjacent strongly pinned vortices. These results can be important for switching applications involving F/S interfaces.

## Acknowledgements
This research has been supported by Council of Scientific and Industrial Research and Department of Science and technology (DST) Government of India. RCB acknowledges J. C. Bose National fellowship of the DST.## Figure Captions:
FIG. 1: (a) X-ray diffraction pattern of the FSB sample. (b) A sketch of the basal plane of the different unit cells of FSB on MgO, with CFS plane rotated $45^o$ with respect to NbN and MgO planes. (c) R vs. T measurements for pure NbN and FSB samples. (d) M vs. T measurements in both FC (Field Cooled) and ZFC (Zero Field Cooled) mode for the same samples.

FIG. 2: $H_{c2}$ (and H*) versus T as extracted from the R(T) (and M(T)) measurements for both pure NbN and FSB samples. Solid lines through $H_{c2}(T)$ are a fit to the data and H*(T) are a guide to the eye.

FIG. 3: FIG. 3: M-H loops measured for: (a) pure NbN sample with in-plane field at 4 and 9 K; (b) FSB sample with in-plane field at 4 K and in the inset at 12 K; and (c) FSB sample with out-plane field at 4 K and in the inset at 20 K.

FIG. 4: (a) M-H loops measured at different temperatures for the FSB sample. Inset shows the magnified view. (b) Switching fields $H_1$ and $H_2$ versus T extracted from the M-H loops. Inset shows the paramagnetic hysteresis loop width $\Delta M$ versus T at different constant fields.

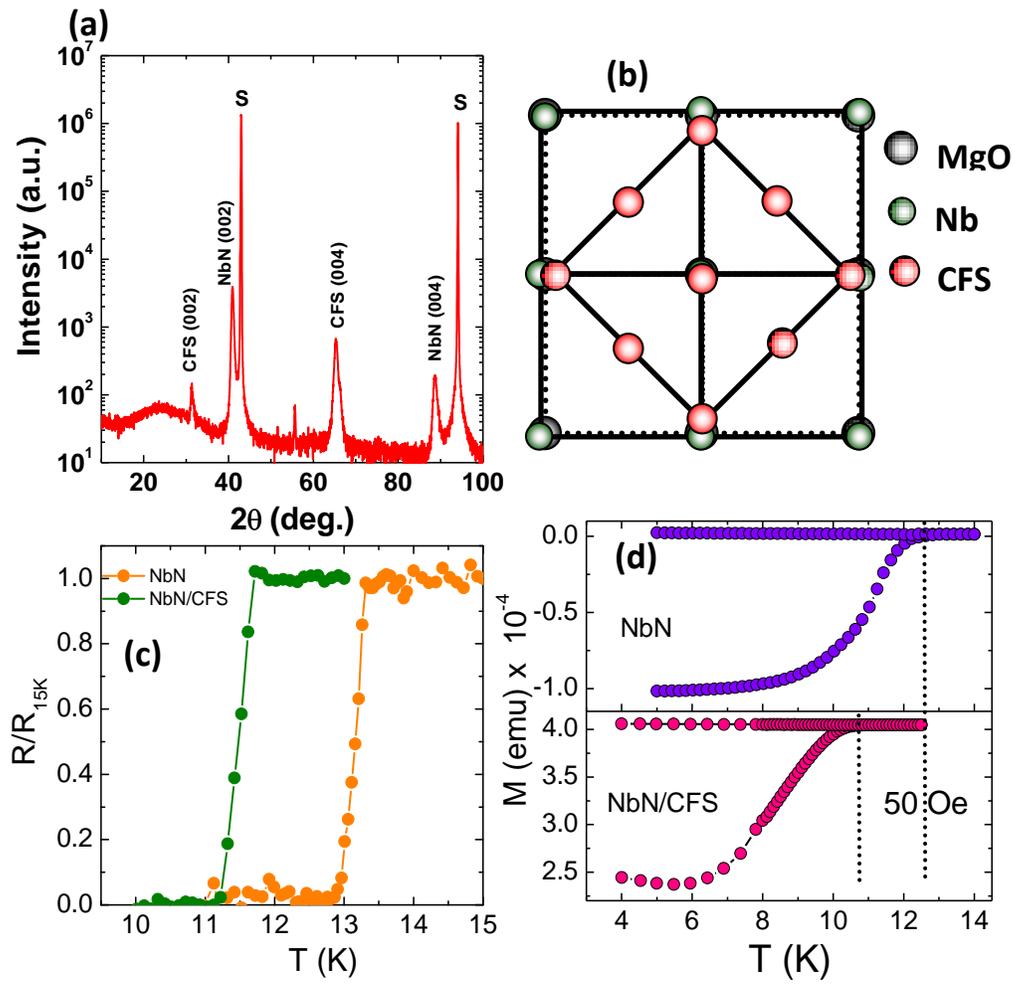

FIG. 1

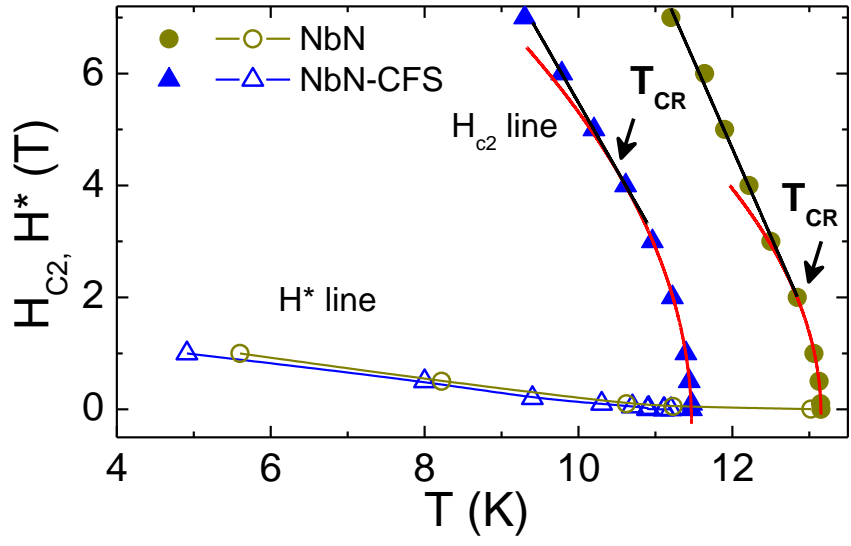

**FIG. 2**

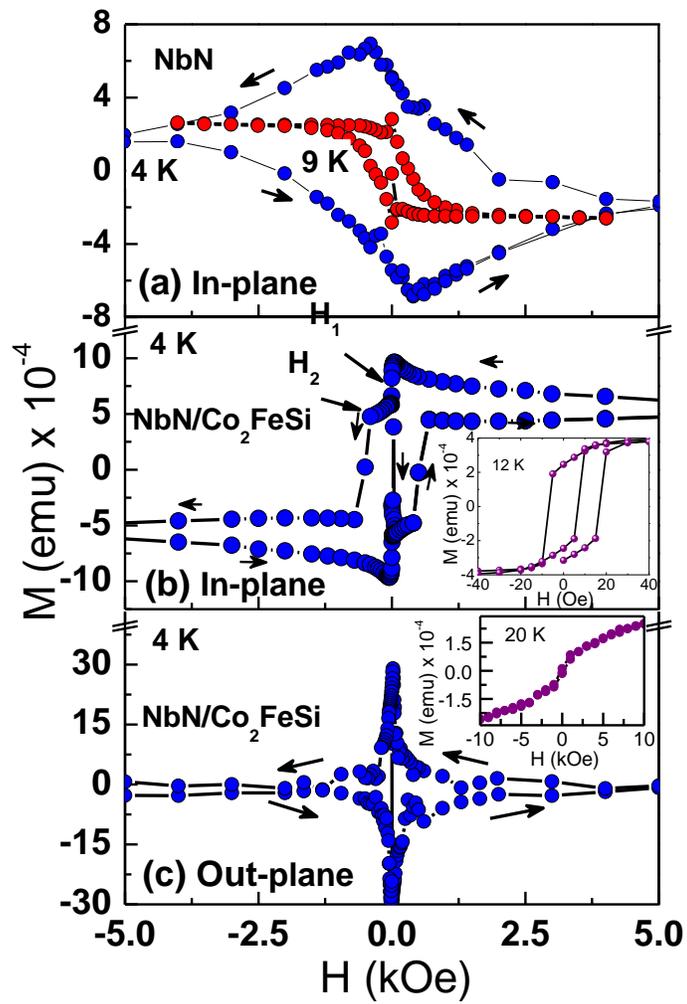

**FIG. 3**

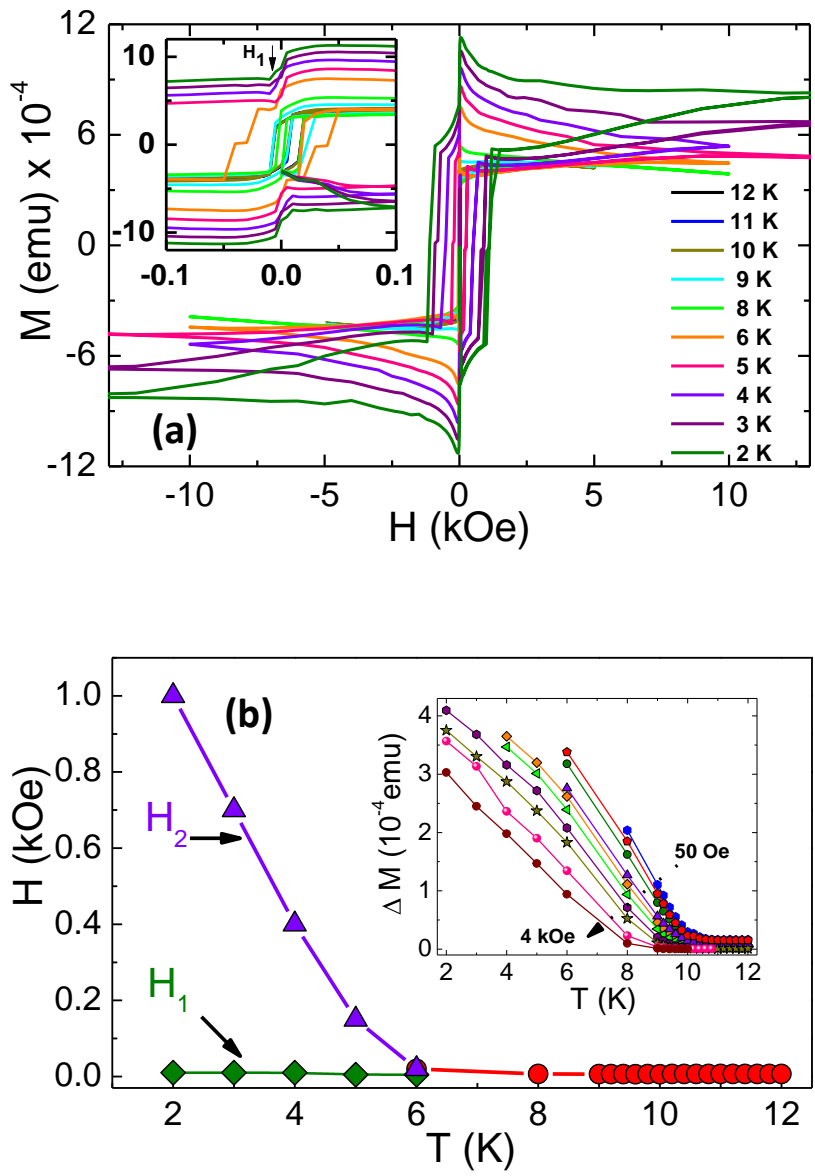

FIG. 4